\documentclass[aps,twocolumn,pra,superscriptaddress,letterpaper]{revtex4-1}
\usepackage{amssymb}
\usepackage{mathtools}
\usepackage[pdftex]{graphicx}
\usepackage[usenames, dvipsnames, svgnames, table]{xcolor}
\usepackage[colorlinks=true, citecolor=Blue,
            linkcolor=BrickRed, urlcolor=ForestGreen]{hyperref}
\usepackage{txfonts}
\usepackage{mathrsfs}
\usepackage{bm}
\usepackage{multirow}
\usepackage{color}
\usepackage{comment}

\begin{document}


\newcommand{\be}{\begin{equation}}
\newcommand{\ee}{\end{equation}}
\newcommand{\R}[1]{\textcolor{red}{#1}}
\newcommand{\B}[1]{\textcolor{blue}{#1}}
\newcommand{\fixme}[1]{\textcolor{orange}{#1}}


\title{Scalable Simulation of Quantum Measurement Process with Quantum Computers}

\author{Meng-Jun Hu} 
\email{humj@baqis.ac.cn}
\affiliation{Beijing Academy of Quantum Information Sciences, Beijing, 100193, China}

\author{Yanbei Chen}
\email{yanbei@caltech.edu}
\affiliation{Theoretical Astrophysics 350-17, California Institute of Technology, Pasadena, CA 91125, USA}
\author{Yiqiu Ma}
\affiliation{Center for gravitational experiment, School of Physics,
Huazhong University of Science and Technology, Wuhan, 430074, Hubei, China}
\author{Xiang Li}
\affiliation{Theoretical Astrophysics 350-17, California Institute of Technology, Pasadena, CA 91125, USA}
\author{Yubao Liu}
\affiliation{Center for gravitational experiment, School of Physics,
Huazhong University of Science and Technology, Wuhan, 430074, Hubei, China}
\author{Yong-Sheng Zhang}
\email{yshzhang@ustc.edu.cn}
\affiliation{Laboratory of Quantum Information, University of Science and Technology of China, Hefei 230026, China }%
\affiliation{Synergetic Innovation Center of Quantum Information and Quantum Physics,
University of Science and Technology of China, Hefei 230026, China}
\author{Haixing Miao} 
\email{haixing@tsinghua.edu.cn}
\affiliation{State Key Laboratory of Low Dimensional Quantum Physics, Department of Physics, Tsinghua University, Beijing, China}


\begin{abstract}
Recent development in quantum information sciences and technologies, especially building programmable quantum computers, provide us new opportunities to study fundamental aspects of quantum mechanics. We propose qubit models to emulate the quantum measurement process, in which the quantum information of a qubit is mapped to 
a collection of qubits acting as the measurement device. One model is motivated by single-photon detection and the other by spin measurement. Both models are scalable to generate Schr\"{o}dinger cat-like state, and their corresponding quantum circuits are 
shown explicitly. Large-scale
simulations could be realised in near-term quantum computers, while classical computers cannot perform the same task efficiently. Due to the scalability of the models, such simulations can help explore the quantum-to-classical boundary, if exists, in the quantum measurement problem. Besides, our protocol to generate cat states may have important applications in quantum computing and metrology.
\end{abstract}

\maketitle


\section{Introduction}
\label{sec:intro}

The quantum measurement problem is at the heart of the foundations of quantum theory, which has not been fully tackled\,\cite{review1, review2, review3}. 
The difficult nature of the measurement problem is mainly due to the linearity of quantum theory that admits the superposition of quantum states, which conflicts with our everyday reality\,\cite{book, zurek2009}. The measurement of a quantum system in the superposition state will result in different outcomes with statistics determined by the Born rule \cite{Dirac}. This inherently irreversible probabilistic cannot be explained by the unitary evolution of a quantum system according to the Schr\"{o}dinger's equation \cite{Wigner}. Various interpretations have been proposed to reconcile the 
issue. The well-known ones include the Copenhagen interpretation\,\cite{Bohr, Hisenberg}, the many-world interpretation \cite{Everett, Wheeler, Dewitt}, the De Broglie-Bohm theory \cite{Bohm1, Bohm2, Bohm3}, and the decoherence theory \cite{zurek1, zurek2, zurek3}. Apart from these interpretations, many specific dynamical models have also been proposed and investigated in detail for elucidating the quantum measurement \cite{review3}, e.g., the von Neumann model\,\cite{von}, quantum statistical models\,\cite{statis1, statis2, statis3}, and system-pointer-bath models \cite{spb1, spb2, spb3}. 

There has been a recurring interest in the measurement topic due to the rapid development of quantum information science and technology in the past several decades\,\cite{qis1, qis2, qis3, qis4, qis5, qis6, qis7, qis8, qis9, qis10}. Significant progress has been made in the manipulation of qubits in various physical systems, e.g., superconducting circuits\,\cite{google, ustc1, ustc2}, trapped ions\,\cite{ion1, ion2, ion3}, atoms in optical lattice\,\cite{atom1, atom2, atom3}, and quantum optics/electronics systems\,\cite{so1,so3,so2}. Advances in quantum technology imply that simulating quantum dynamics with quantum computers has become feasible nowadays\,\cite{Feynman, simulation, simulation2}. 
Since classical reality in quantum measurement is believed to emerge from the building blocks of quantum, the microscopic system plus macroscopic measurement apparatus must be described quantum mechanically according to mainstream measurement theory\,\cite{review1, review2, review3}. 
It is thus natural to ask whether and how quantum computers can be used to simulate and investigate the quantum measurement process. 
While general-purpose and universal quantum computers are still far away, current noisy intermediate-scale quantum (NISQ) processors and more advanced ones in the near future have already provided this opportunity\,\cite{NISQ}.

\begin{figure*}[tp]
\includegraphics[scale=0.65]{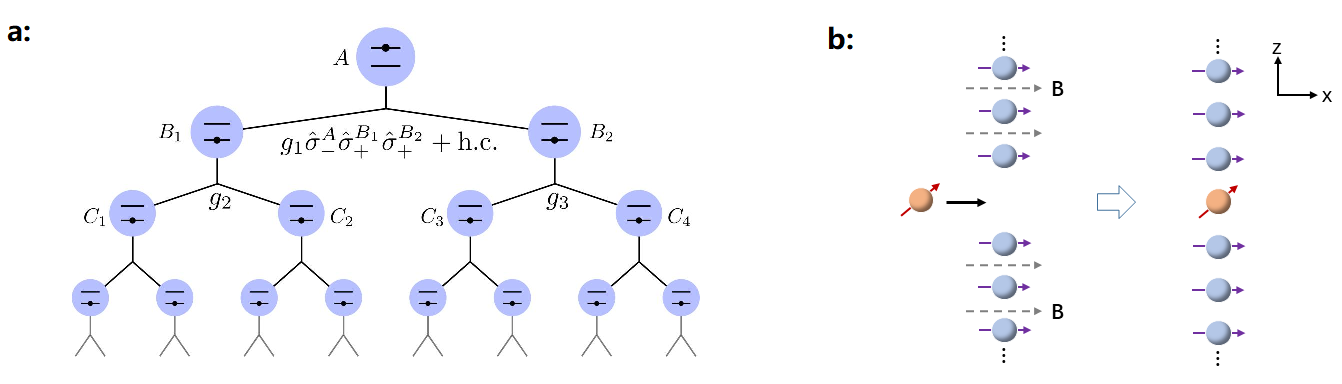}
\caption{Schematic diagram of two qubit measurement models. {\bf a:} Qubit tree network model for simulating
the single-photon detection. Each qubit is coupled to two qubits in the next layer via a three-qubit interaction. The qubit in the first layer is the system qubit, its excitation representing the photon; Qubits in the last layer of the network are measurement qubits, their excitations representing the electrons emerged from the photodetector. {\bf b:} Spin measurement model, in which the system spin (qubit) interacts with the measurement spins (qubits) via an Ising-type Hamiltonian. All measurement spins are prepared in the superposition state $|+\rangle$, e.g., by applying a ``magnetic field" $B$ along the x-direction, and the field will be turned off when the measurement process begins. }
\label{fig:model}
\end{figure*} 

In order to simulate with quantum computers, qubit models of the measurement process are necessary.
Here, we propose two dynamical models with a scalable number of qubits. The system to be measured is a single qubit, and the measurement device is modeled as a collection of qubits.
The first model is motivated by the single-photon detection in quantum optics, in which the system qubit is de-excited after the measurement. The second model is related to the non-destructive spin measurement, and the system qubit can be used for subsequent measurements. In both models, if the system qubit is in the definite state $|0\rangle_{s}$ or $|1\rangle_{s}$, the corresponding final state of the measurement qubits will be in a collective state either $|\bm 0\rangle_{m}$ or $|\bm 1\rangle_{m}$\,\footnote{It should be noted that the collective state, e.g., $|\bm 0\rangle_{m}$, will only imply that most of qubits in the state $|0\rangle$ when considering the physical realisation of the models.}.
Given an arbitrary superposition state of the system qubit: $\alpha|0\rangle_{s}+\beta|1\rangle_{s}$, the first model leads to the final state of the compound system as 
\begin{equation}\label{eq:psi_1}
|\Psi(t_f)\rangle_{\rm first\,model} =|0\rangle_{s}\otimes\left(\alpha|\bm 0\rangle_{m}+\beta|\bm 1\rangle_{m}\right)\,,
\end{equation}
while the second model results in a cat-like final state 
\begin{equation}\label{eq:psi_2}
|\Psi(t_f)\rangle_{\rm second\,model}=\alpha|0\rangle_{s}|\bm 0\rangle_{m}+\beta|1\rangle_{s}|\bm 1\rangle_{m}\,.
\end{equation}
In both models, the interaction will lead to a proliferation of the quantum information from a single qubit to a collection of qubits---a prototype of the measurement process. 
Since the models are in the paradigm of unitary evolution, the purpose of this work is not to solve the measurement problem but to explore the prospect of using quantum computers to study the emergence of classical reality in the measurement process. 

\section{Description of Qubit Measurement Models}
\label{sec:model}

The first model originates from the phenomenology of the single-photon detection in quantum optics. In the photodetector, a single photon leads to the excitation of many electrons which creates a macroscopic observable current. We model the continuous translational degrees of freedom of electrons by discrete two-level systems, i.e., qubits. 
The model is illustrated in 
Fig. 1a. It is a tree 
network of qubits, and they are interacting
with each other via a three-qubit 
interaction, i.e., each
qubit is coupled to two 
qubits in the next layer.
The first qubit, or the 
system qubit, is viewed as the 
single photon (or the atom that generates
the single photon). The last layer of the 
network
mimics the 
electrons that generate the observed current of the photodetector. 

The Hamiltonian 
of the entire network reads 
\begin{align}\nonumber
\hat H=&\sum_i\hbar \frac{\omega_i}{2}\hat \sigma^{i}_Z+\hbar g_1\hat \sigma_-^{A}\hat\sigma_+^{B_1}\hat\sigma_+^{B_2} \\&+ \hbar g_2\hat\sigma_-^{B_1}\hat\sigma_+^{C_1}\hat\sigma_+^{C_2}
 + \hbar g_3\hat\sigma_-^{B_2}\hat\sigma_+^{C_3}\hat\sigma_+^{C_4} + \cdots + {\rm h.c.}\,.
\end{align}
Here $\omega_{i}$ is the transition
frequency of each qubit, $\hat\sigma_{\pm}\equiv (\hat\sigma_{X}\pm\hat\sigma_{Y})/2$ are ladder operators defined by Pauli operators $\hat \sigma_X$ and $\hat \sigma_Y$; $g_{1, 2, 3...}$ represent the coupling strength, and ${\rm h.c.}$ means Hermitian conjugate. Each term in the Hamiltonian describes the process 
of de-exciting one qubit and exciting two qubits in the next layer, while the conjugate part describes the process of de-exciting two qubits and exciting one qubit in the previous layer. In this model, the total number 
of qubits equals to
$N_{\rm qubits}= 2^{N_{\rm layers}} -1\,$
with $N_{\rm layers}$ being the number of 
layers. As $N_{\rm qubits}$ increases, the dimension of Hilbert space of qubits increase exponentially, and thus it is impossible to simulate such model efficiently with classical computers. 

The dynamical evolution depends on 
the coupling strength $g_i$. We control 
them such that if qubit $A$ is in the excited state $|1\rangle$ initially, the qubits in the last layer will end up in a collective excitation state $|\bm 1\rangle$. The other qubits in the network will go through 
some intermediate state and be 
in the ground state at the end of the
simulation. The simplest realisation, 
which can also be implemented by 
the quantum circuit shown later, 
is to make $g_i$ 
being pulse functions with the same 
amplitude of $g$ and the same duration
of $\tau=\pi/(2g)$. In this case, 
only two layers are turned on for 
each period of
$\tau$ when the qubits involved
undergo a complete state transfer. 
Take the example of three layers 
with 7 qubits, we have 
\begin{align}\nonumber
    g_1(t) & = g\,\left[\Theta(t)-\Theta(t-\tau)
    \right]\,,\\
    g_2(t) & = g_3(t)=g\,\left[
    \Theta(t-\tau) -\Theta(t-2\tau)\right]\,
\end{align}
with $\Theta(t)$
the Heaviside function. If the 
initial state of the network is 
\begin{equation}
|\Psi(t_0)\rangle = (\alpha|0\rangle_s +\beta|1\rangle)_s)\otimes |\bm 0\rangle_m\,. \end{equation}
When the simulation 
ends at $t_f = 2\tau$, we will obtain 
the final state shown in Eq.\,\eqref{eq:psi_1}. 

The second model, as shown in Fig. 1b, describes the spin measurement, which is motivated by Stern-Gerlach experiment. We use 
a collection of $2N$ 
spins as the 
measurement device to probe 
the state of the system 
spin directly, in which they interact with the system spin 
$\hat \sigma_Z^{s}$ via Ising-type Hamiltonian: 
\begin{equation}
\hat{H}(t)=-\sum_{sm}J_{sm}(t)\hat\sigma_{Z}^{s}\hat \sigma_{Z}^{m}-\sum_{mn\neq s}J_{mn}\hat\sigma_{Z}^{m}\hat\sigma_{Z}^{n}-h(t)\sum_{m}\hat\sigma_{X}^{m}\,,
\end{equation}
where $h(t)$ represents
the transverse field $B$ applied 
to the measurement spins. 
\begin{figure*}[!t]
\includegraphics[scale=0.45]{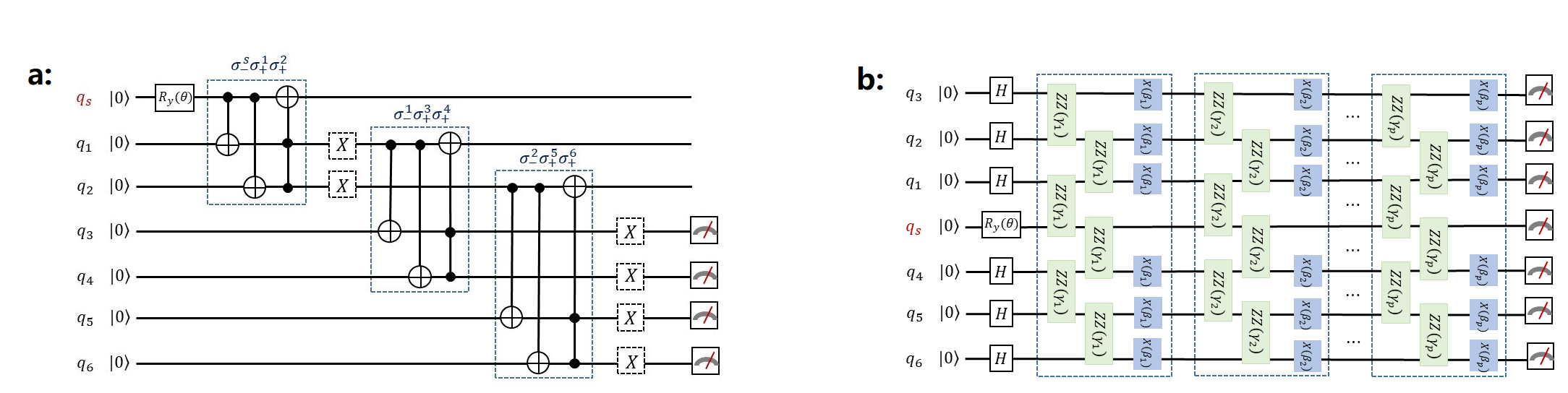}
\caption{Quantum circuits for two qubit-measurement models. {\bf a:} tree network model for three layers with 7 qubits. Gate operations within the blue dotted line realize the three-qubit interaction by 
using two CNOT gates and one Toffoli gate.
The additional $X$ gate represents a small error that causes qubit flip to mimics the dark count/loss in the photodetector. {\bf b:} the spin measurement model. Here we show the 7-qubit case with the nearest-neighbour interaction. The gate operations in dotted blue box are identical except with different parameters $\gamma_{i}, \beta_{i}$. Light green box represents two-qubit $ZZ$ gate $ZZ(\gamma)\equiv e^{i\gamma\hat \sigma_{Z}\hat \sigma_{Z}}$, and light blue box represents single-qubit rotation $X$ gate $X(\beta)\equiv e^{i\beta\hat \sigma_{X}}$. More qubits circuit of two models can be similarly obtained.}
\label{fig:circuit}
\end{figure*}

We apply a strong 
field at $t<t_0$: $h(t)=h_0\Theta(t_0-t)$
to prepare the state of the 
measurement spins in $|+\rangle\equiv (|0\rangle+|1\rangle)/\sqrt{2}$ with $|0\rangle$ and $|1\rangle$ representing spin-up and spin-down, respectively.
The 
system spin is prepared in a superposition state: $\alpha|0\rangle+\beta|1\rangle$. 
The resulting 
state of the entire 
system is thus given by
\begin{equation}\label{eq:psi_t0}
|\Psi(t_{0})\rangle=(\alpha|0\rangle+\beta|1\rangle)\otimes|+\rangle^{\otimes 2N}.
\end{equation}
At $t=t_0$, the transverse field is turned off and we turn on the 
interaction between the system 
spin and the measurement spins: $J_{sl}(t)=J_{sl}>0$, and the Hamiltonian becomes zero-field Ising Hamiltonian
\begin{equation}\hat{H}(t\ge t_0)= \hat{H}_{\rm Ising}= -\sum_{ij}\,J_{ij}\hat \sigma_{Z}^{i}\hat\sigma_{Z}^{j}\,.
\end{equation}

The ground state of the
above Hamiltonian is double degenerate, i.e., $| 0, 0, ..., 0\rangle$ and $|1, 1, ..., 1\rangle$.
Therefore, if we can drive 
the entire system into its 
ground state, we will obtain 
the final state shown in Eq.\,\eqref{eq:psi_2}. However, the  dynamical evolution of spins system
under the Ising Hamiltonian does not leads to the ground state. This is because 
the initial state in Eq.\,\eqref{eq:psi_t0} can be recast as 
\begin{equation}
|\Psi(t_{0})\rangle = \frac{1}{2^{N}}[|E_{0}\rangle+\sum_{i}|E^{i}_{1}\rangle+\sum_{m, n}|E^{m,n}_{2}\rangle+\cdots+|E_{N}\rangle],
\end{equation}
where $|E_{0}\rangle\equiv \alpha|0, 0, ..., 0\rangle+\beta|1, 1, ..., 1\rangle$ is the ground state of $\hat{H}_{\rm Ising}$, and $|E_{i}\rangle$ represent the degenerate excited states . The dynamical evolution under $\hat{H}_{\rm Ising}$ only introduces relative phase for different eigenstates. In order to obtain the desired final ground state $|E_{0}\rangle$, we introduce a control field
acting on the measurement
spins. Here we adopt Dicke-type Hamiltonian 
 in the mean field 
approximation\,\cite{Dicke1, Dicke2}: 
\begin{equation}
\hat{H}_{\rm Dicke} =\hbar\, \phi(t)\sum_{k=1}^{2N}\hat \sigma_{X}^{k}\,.
\end{equation}
We want to design the time-dependent 
control field $\phi(t)$ 
such that the coefficients of the excited states $|E_{N}\rangle$ decay away and only the lower energy states, e.g, $|E_{0}\rangle$ will remain by the end
of evolution. In the following, we show that the target evolution can be well approximated with a novel variational quantum circuit.

\begin{figure*}[tp]
\includegraphics[scale=0.42]{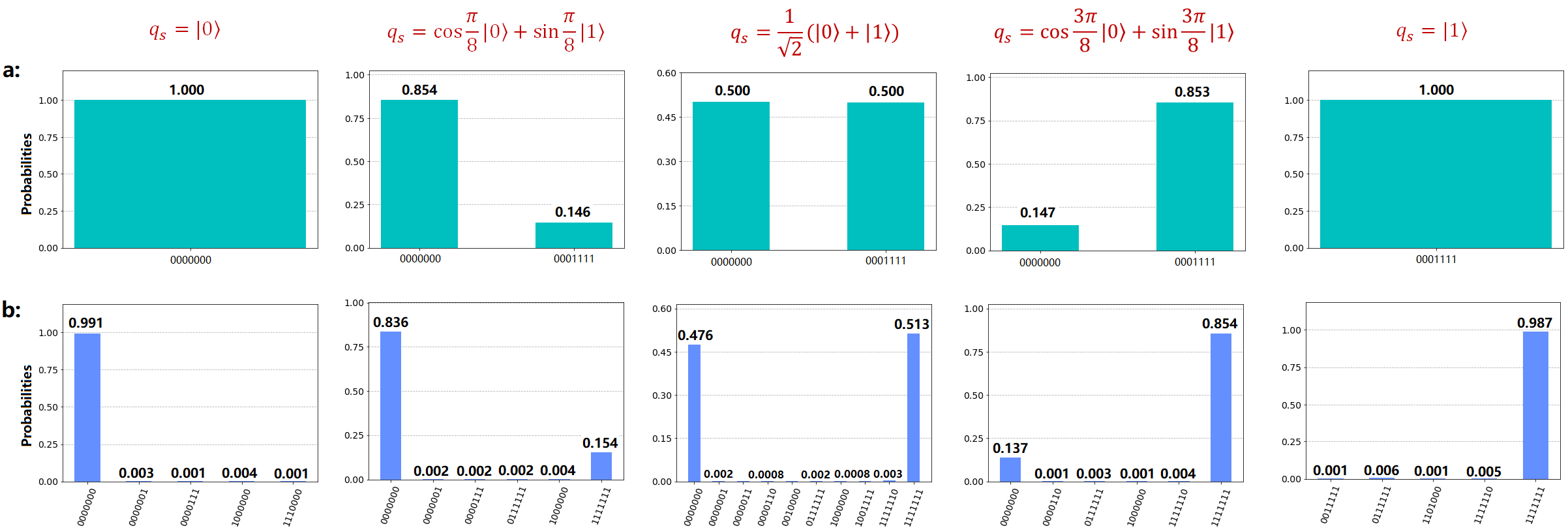}
\caption{Simulation results for {\bf a:}
the tree network model, and {\bf b:}
the spin measurement model (second row) for 7 qubits. The system qubit is 
prepared in different initial states shown on the top. The bar diagram indicates 
the probability of 
different final states (indicated by the labels on the horizontal axis) for all 
the qubits. The optimal parameters of spin measurement circuit are determined by variational optimization with $q_{s}=|0\rangle$. Once optimal parameters are obtained, we fixed them in quantum circuit and run simulation for other states of $q_{s}$. }
\label{fig:result}
\end{figure*} 

\section{Quantum circuits of two models}
The key to realize the above two qubit measurement models with quantum computers is constructing the corresponding quantum circuits, which are shown in  Fig.\,\ref{fig:circuit}. We use $7$ qubits as an example and the circuits can be extended to the case with more qubits. In the circuits, $q_{s}$ represent the system qubit and the other qubits are the measurement qubits. 

Fig.\,\ref{fig:circuit}a shows the quantum circuit of $3$ layer qubits tree network model with $7$ qubits. The core module of the circuit is the realization of Hamiltonian term proportional to 
$g \hat \sigma_{-}^i\hat \sigma_{+}^j\hat \sigma_{+}^k$. With the 
pulsed interaction described in 
the previous section, an interaction 
duration of $\tau = \pi/(2g)$ 
results in de-exciting one qubit and exciting two qubits in the next layer. 
Its equivalent gate operation,  shown in Fig. \ref{fig:circuit}a, consists of two CNOT gates and one Toffoli gate. 

Take the process 
involving $\hat \sigma_{-}^{s}\hat \sigma_{+}^{1}\hat \sigma_{+}^{2}$ for instance, two CNOT gates with $q_{s}$  as the qubit are executed sequentially and then Toffoli gate with $q_{1}, q_{2}$ as the control qubits is executed. If $q_{s}$ is in the ground state $|0\rangle$, $q_{1}, q_{2}$ will not be excited and stay in the ground state. If instead $q_{s}$ is in the excited state $|1\rangle$, the gate operation guarantees that $q_{1}, q_{2}$ be excited to $|1\rangle$ and meanwhile $q_{s}$ be de-excited to the ground state $|0\rangle$. 
For the general case that $q_{s}$ is prepared in the superposition state $\alpha|0\rangle_{s}+\beta|1\rangle_{s}$ by using the gate operation $R_{y}(\theta)$, the state of $q_{1}, q_{2}$ after operations be $\alpha|0\rangle_{1}|0\rangle_{2}+\beta|1\rangle_{1}|1\rangle_{2}$. Using a similar analysis, the final state of $q_{3}, q_{4}, q_{5}, q_{6}$ in Fig.\,\ref{fig:circuit}a will be $\alpha|0\rangle_{3}|0\rangle_{4}|0\rangle_{5}|0\rangle_{6}+\beta|1\rangle_{3}|1\rangle_{4}|1\rangle_{5}|1\rangle_{6}$. 

Extending the above analysis to the situation with more layers, the final state of the measurement qubits (the last layer in the network) will indeed be $\alpha|\bm 0\rangle+\beta|\bm 1\rangle$,
i.e., the one shown in Eq.\,\eqref{eq:psi_1}; all the other qubits will be in the ground state $|0\rangle$. The quantum circuit for the tree network model thus demonstrates how the quantum information of the system qubit can be transferred or, in some sense, 
augmented to the measurement qubits. 
Note that the model only emulates the 
ideal single photon detection. In the realistic 
situation, however, there also exists 
dark counts and 
imperfect detections. The former describes clicks of the detector even when there were no signal photon coming in; the latter refers to the situation when the detector does not response to the signal photon. These two phenomena can be accounted for in the quantum circuit if we introduce a probabilistic flip gate operation $X$ after the three-qubit interaction, which 
is also shown in Fig.\,\ref{fig:circuit}a.


For the spin measurement model, 
unlike the previous model, there is 
no specific interaction duration that 
could lead to one meaningful state 
transition. Here we propose to approximate 
the dynamical evolution by implementing the idea of variational quantum circuit\,\cite{VQA}. The circuit evolution is given by
\begin{equation}
\hat{U}|\Psi(t_{0})\rangle=\hat{U}(\beta_{p})\hat{U}(\gamma_{p})\cdots\hat{U}(\beta_{1})\hat{U}(\gamma_{1})|\Psi(t_{0})\rangle\,.
\end{equation}
Here $\hat{U}(\gamma)=e^{i\gamma\sum_{<ij>}\hat{\sigma}^{i}_{Z}\hat{\sigma}^{j}_{Z}}=\prod_{<ij>} e^{i\gamma\hat {\sigma}_{Z}^{i}\hat{\sigma}_{Z}^{j}}$ and $\hat{U}(\beta)=e^{i\beta\sum_{k=1}^{2N}\hat{\sigma}^{k}_{X}}=\prod_{k=1}^{2N}e^{i\beta\hat{\sigma}^{k}_{X}}$, which are related to the  evolution under $\hat{H}_{\rm Ising}$ and $\hat{H}_{\rm Dicke}$, respectively. It is interesting to note that this variational quantum circuit is very similar to the circuit of quantum approximate optimization algorithm (QAOA) \cite{QAOA} except that there is no single qubit rotation $e^{i\beta\hat{\sigma}_{X}}$ on the system qubit $q_{s}$. We limit 
ourselves to the case that $\hat{H}_{\rm Ising}$ only takes the nearest-neighbor interaction. In Fig.\,\ref{fig:circuit}b, we show the corresponding circuit for 7 qubits. The Hardmard gates on the measurement qubits $q_{1}\cdots q_{6}$ and the single qubit rotation gate on the system qubit $q_{s}$ prepares the initial state shown in Eq.\,\eqref{eq:psi_t0}. Quantum operations within the blue dotted line realizes the evolution $\hat{U}(\beta)\hat{U}(\gamma)$, and it is repeated $p$ times in the circuit.

To obtain the desired target final state shown in Eq.\,\eqref{eq:psi_2}, we need
to specify the $2p$ parameters $\lbrace\vec{\gamma}, \vec{\beta}\rbrace$
for given circuit depth $p$. Since 
our target is to obtain the ground state of the 
Ising model, we adopt the idea of variational quantum algorithm in which optimal parameters $\lbrace\vec{\gamma}, \vec{\beta}\rbrace$ are determined by using classical optimizer that minimizes the expectation value of Ising Hamiltonian\,\cite{QAOA, VQA, cover, Qcover}. Specifically, we prepare the parameterized quantum circuit with a finite $p$ as shown in Fig.\,\ref{fig:circuit}b, and then obtain the optimized parameters $\lbrace\vec{\gamma}^{*}, \vec{\beta}^{*}\rbrace$ using a classical optimizer, e.g., COBYLA to minimize the expectation value $\langle\Psi(\vec{\gamma}, \vec{\beta})|\hat{H}_{\rm Ising}|\Psi(\vec{\gamma}, \vec{\beta})\rangle$. Finally, the optimal parameters $\lbrace\vec{\gamma}^{*}, \vec{\beta}^{*})$ are fixed in the model circuit for the final sampling with different input state $q_{s}$.  

\section{Simulation Results}

Here we present and discuss the corresponding simulation results. We 
simulate the corresponding quantum circuits of two models by using IBM {\it qiskit} package \cite{qiskit}. Fig.\,\ref{fig:result}a shows the result 
for the $7$ qubits tree network without introducing the qubit flip error $X$. It matches the theoretical prediction of the final state shown in Eq.\,\eqref{eq:psi_1} for different initial states of the system qubit $q_{s}$. In the Appendix, we also present the results with different flip error probability to simulate dark counts and imperfect detection. Interestingly, 
those imperfections do not break 
the mirror symmetry between the 
final probability distributions for 
two different initial states $\alpha|0\rangle+\beta|1\rangle$ and $\beta|0\rangle+\alpha|1\rangle$, respectively. 

For the spin measurement model, we use the variational quantum circuit with a finite circuit depth $p$.  In general, the larger $p$ is, the better the simulation results are. However, more time will be needed to optimize $2p$ parameters $\lbrace\vec{\gamma}, \vec{\beta}\rbrace$. In practice the performance of quantum computers will also limit the depth that can be efficiently implemented. For $7$ qubits, we choose $p=3$ that leads to a good performance. More qubits results with $p=12$ are shown in Appendix. 
To determine the optimal $2p$ parameters $\lbrace\vec{\gamma}^{*}, \vec{\beta}^{*}\rbrace$, 
we execute the parameterized quantum circuit with $2p$ initial guess for these 
parameters and obtain a final state $|\Psi(\vec{\gamma}, \vec{\beta})\rangle$; we calculate the expectation value $\langle\Psi(\vec{\gamma}, \vec{\beta})|\hat{H}_{\rm Ising}|\Psi(\vec{\gamma}, \vec{\beta})\rangle$. The classical optimizer, e.g., COBYLA, is called to update parameters to minimize the above 
expectation value until the condition of convergence is satisfied. Even though 
the optimal values would depend on 
the initial state of the system qubit, 
it is interesting that the dependence is not strong. As shown in Fig.\,\ref{fig:result}b, we have optimised for the initial state of $q_s$ being $|0\rangle$ and obtained a final state very close to the desired one $|0,0,0,0,0,0,0\rangle$. The results for  other input initial states still match well with the expected outcome.
Therefore, after the optimal parameters $(\vec{\gamma}^{*}, \vec{\beta}^*)$ are found 
for one specific initial state, we can perform measurement of $q_{s}$ with arbitrary unknown quantum state $\alpha|0\rangle+\beta|1\rangle$, which 
is ideal for the physical realisation. 

\section{Discussion and Conclusion}
\label{sec:conclusion}

The quantum measurement problem is one of the most fundamental issues in quantum theory. The rapid development of quantum information science in recent years has already provided many insights into solving the measurement problem \cite{qis3, qis4, qis5, qis6, qis7}. The purpose of this work is to motivate the simulations of the measurement process with scalable and programmable quantum computers. Since the randomness in measurement outcomes cannot be explained within the current framework of quantum theory, it should be noted that quantum simulations based upon 
linear and unitary evolution alone cannot solve the problem. The quantum computers, however, can help explore the possible quantum-to-classical boundary in the quantum measurement process. In particular, the scalability of quantum computers implies that we can study complex dynamical characteristics of the measurement process when more and more qubits are included. Such a platform could also verify various measurement theories, e.g., quantum Darwinism which emphasizes the importance of redundancy in environment for the emergence of classical objective reality\,\cite{qis9}. Besides, the impossibility to perform a large scale simulation with many qubits offer a new way to demonstrate quantum advantages in the NISQ era.

In conclusion, we have proposed two qubit measurement models. The qubit tree network model, which is inspired by the single photon detection in quantum optics, describes how the state information of a single system qubit can be propagated and amplified to the last layer of network that consists of many qubits. The state 
of the system qubit is destroyed after the measurement process in this model. In contrast, the spin measurement model describes how the information of the system qubit is mapped to to the measurement qubits without being destroyed. So far, we have been focusing on the circuit
models for gate-based quantum computers. It is, however, also possible to consider analog quantum computers or simulators \cite{A1, A2, A3, A4, A5}. Taking Rydberg atom-based programmable quantum simulator \cite{A2, A3} as example, we may arrange atoms in the tree network configuration, and take advantage of the blockade effect to realize the three-qubit interaction. Similarly, due to the natural description of ion interaction with Ising Hamiltonian, it may be more suitable to consider ion trap-based quantum simulator \cite{A4, A5} for the spin measurement model. Therefore, we can perform simulations of these two measurement models in both gate-based and analog quantum computers to explore 
the limit of quantum realm. 

\hfill

\begin{acknowledgments}
We would like to thank Huan Zhang, Wei-Feng Zhuang, Dong E. Liu and 
the MQM group for valuable discussions. M. H. is supported by Beijing Academy of Quantum Information Sciences. Y. C. and X. L. have been funded by the US National Science Foundation, and the Simons Foundation. Y. M. and Y. L. are supported by the start-up fund provided by Huazhong University of Science and Technology. Y. Z. is supported by the National Natural Science Foundation of China (Grant No. 92065113) and Anhui Initiative in Quantum Information Technologies.
H. M. is supported by State Key Laboratory of Low Dimensional Quantum Physics and the start-up fund from Tsinghua University.
\end{acknowledgments}




\appendix

\begin{figure*}[tp]
\includegraphics[scale=0.35]{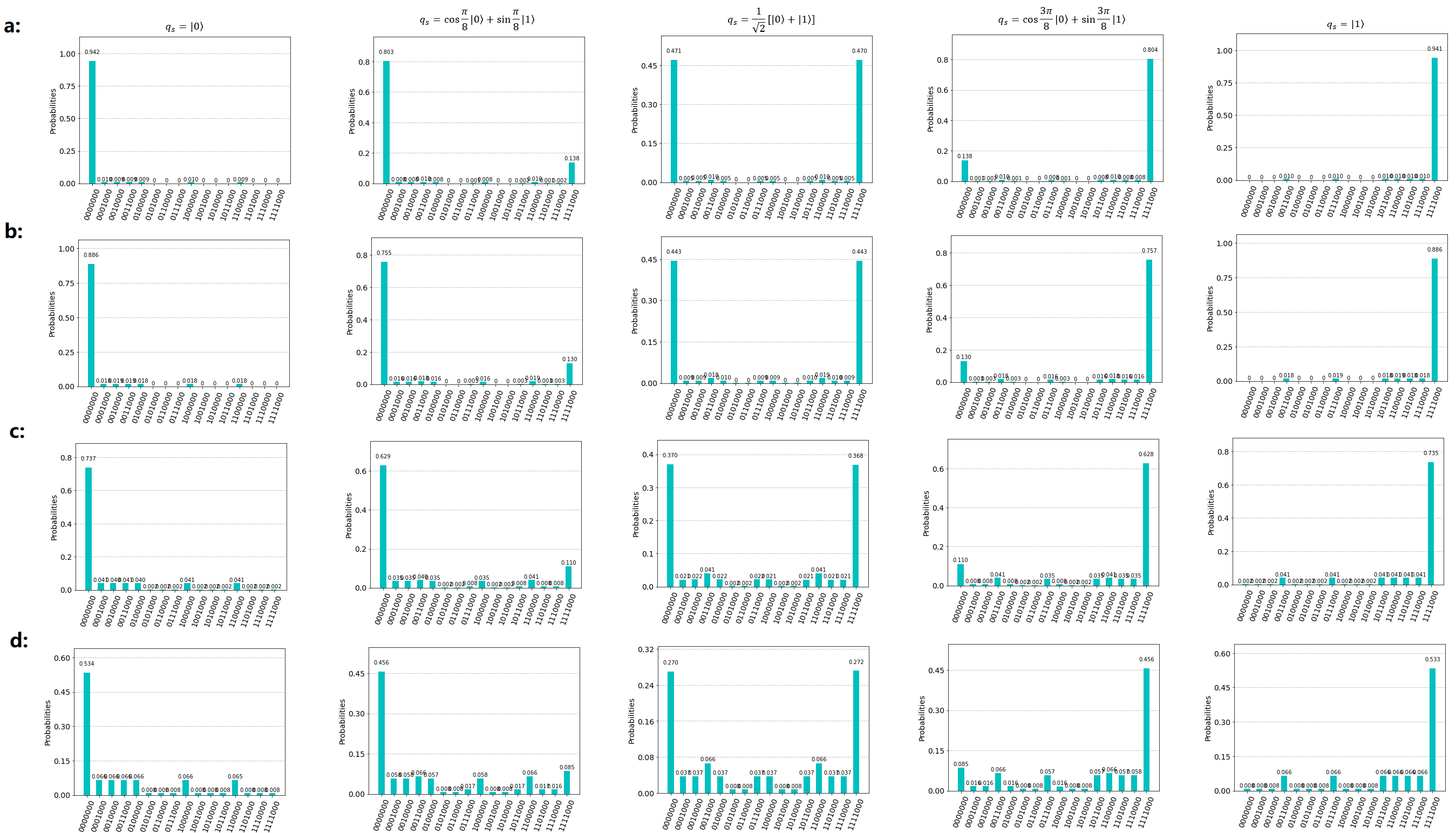}
\caption{Qubit tree network model with flip error. Results of different flip probability $P$ of qubits tree network circuit with {\bf a:} P=0.01, {\bf b:} P=0.02, {\bf c:} P=0.05, {\bf d:} P=0.1.}
\end{figure*}

\begin{figure*}[tp]
\includegraphics[scale=0.4]{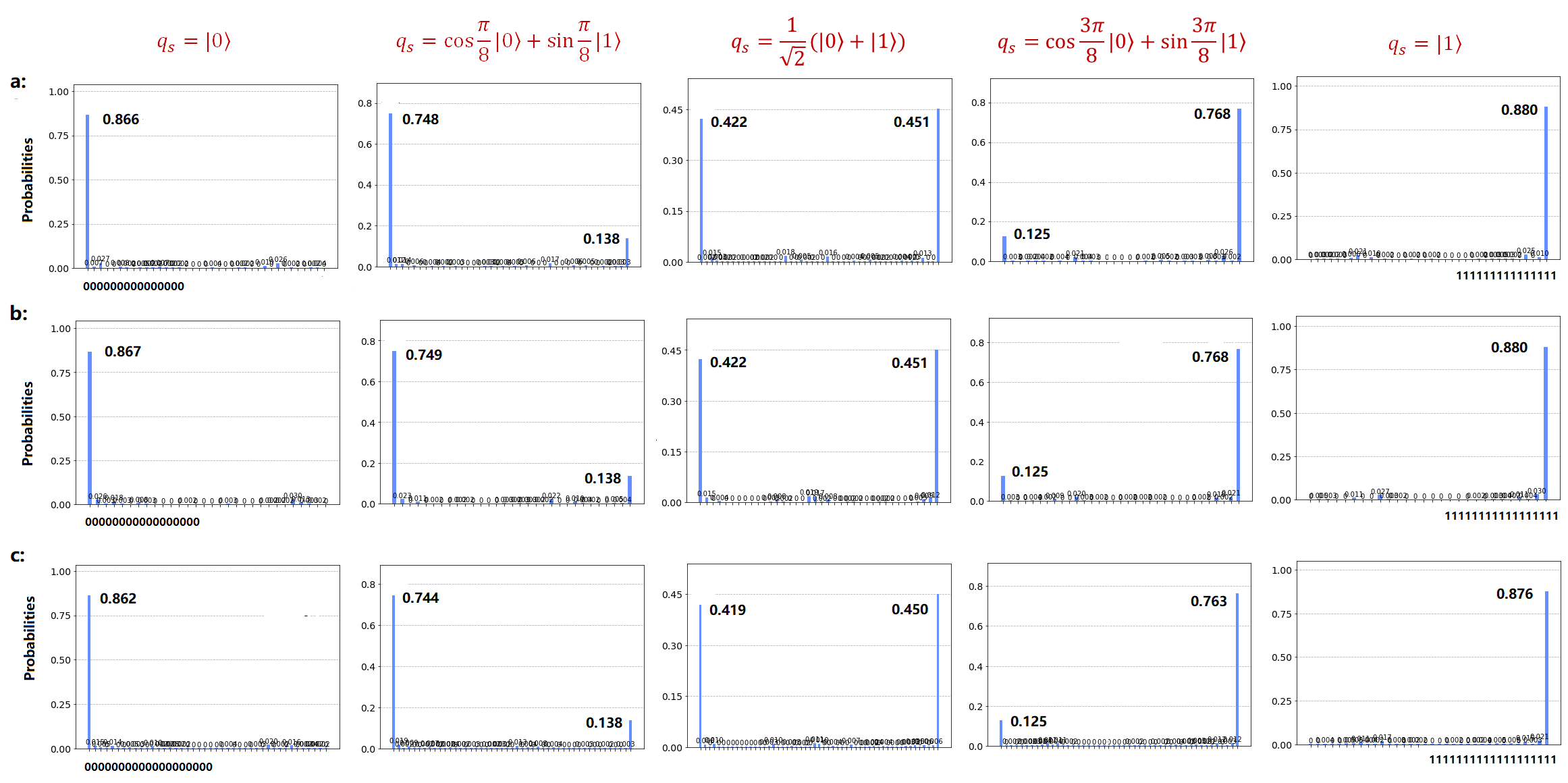}
\caption{Variational simulation of spin measurement model with different qubits number and fixed circuit layer of $p=12$. Results of different qubits number {\bf a:} N=15, {\bf b:} N=17, {\bf c:} N=19.}
\end{figure*} 

\section{More Simulation Results}


\end{document}